\documentclass[preprint,aps]{revtex4}
\usepackage{graphicx}
\usepackage{epsfig}
\usepackage{CJK}
\usepackage{latexsym,amssymb,amsmath}
\usepackage{amsmath}
\usepackage{supertabular}
\usepackage{textcomp}
\usepackage{cases}

\begin{document}

\title{Corrections induced by the GUP to the Lamb shift of an accelerated atom interacting with a quantum scalar field}
\author{Zhi Wang$^{1,2}$$\footnote{E-mail address: zwangphys@163.com}$, Yi Yang$^{1,2}$$\footnote{E-mail address: yiyang@mail.gufe.edu.cn}$, Zhengwen Long$^{3}$$\footnote{E-mail address: zwlong@gzu.edu.cn}$}
\affiliation{$^{1}$School of Mathematics and Statistics, Guizhou University of Finance and Economics, Guiyang 550025, China\\
	$^{2}$School of Big Data Statistics, \\
	Guizhou University of Finance and Economics, Guiyang 550025, China\\
	$^{3}$School of Physics, Guizhou University, Guiyang 550025, China}

\begin{abstract}

We investigate the effect of the GUP on the Lamb shift of a two-level atom interacting with a real massless scalar quantum field, within the DDC formalism. For an atom undergoing inertial motion, uniform acceleration, and uniform circular motion, we analyze the separate contributions of vacuum fluctuations and radiation reaction.
We first derive the statistical functions of the field along the atom's trajectories for the three types of motion, expressing them as frequency integrals, and then employ them to calculate the vacuum fluctuation and radiation reaction contributions to the radiative level shift.
We show that the GUP-modified Lamb shift of the two-level atom arises entirely from vacuum fluctuations and acquires additional corrections proportional to $\beta$.
We focus in particular on the acceleration-dependent GUP corrections. For a uniformly accelerated atom, the GUP corrections comprise thermal and nonthermal parts. At low accelerations, the thermal part exhibits nonmonotonic behavior, and is proportional to $a^4$ in the limit $a/\omega_0 \to 0$; the nonthermal part, by contrast, grows nonlinearly and monotonically, exceeding the thermal part by nearly two orders of magnitude at large accelerations.
For an atom in uniform circular motion, the GUP corrections are purely nonthermal and also display a nonlinear, monotonic dependence on acceleration, increasing or decreasing steeply according to the sign of $\beta$. For the same $\beta$, the corrections are larger in uniform circular motion than in uniformly accelerated motion, since the former involves terms proportional to both $a^2$ and $a^3$, whereas the latter contains only $a^2$ terms.

\end{abstract}

\maketitle

\section{Introduction}

The radiative properties of atoms, including spontaneous emission, excitation, and the Lamb shift, constitute one of the most fundamental testing grounds of quantum electrodynamics. The Lamb shift, the tiny displacement of atomic energy levels induced by the coupling to the vacuum fluctuations of the quantized electromagnetic field, was first measured by Lamb and Retherford in 1947 \cite{Lamb1947PR} and provided a decisive impetus for the development of renormalized QED \cite{Bethe1947PR}. A key conceptual advance in understanding the physical origin of radiative processes was achieved by Dalibard, Dupont-Roc, and Cohen-Tannoudji (DDC), who proposed a formalism based on the symmetric ordering of atomic and field operators \cite{Dalibard1982JPhys,Dalibard1984JPhys}. In this framework, the rate of change of the mean atomic energy can be unambiguously separated into two Hermitian, physically distinct contributions: that of vacuum fluctuations and that of radiation reaction. For an inertial ground-state atom, these two contributions cancel exactly, ensuring the stability of the vacuum; for an excited-state atom, they add constructively, giving rise to spontaneous emission. This decomposition resolved a long-standing ambiguity originating from the freedom in choosing the operator ordering in the Heisenberg-picture interaction Hamiltonian \cite{Milonni1976PhysRep,Milonni1994Book}.

The DDC formalism was subsequently extended beyond inertial atoms in flat spacetime. Audretsch and Müller \cite{Audretsch1994PRA,Audretsch1995PRA} applied the DDC approach to a uniformly accelerated two-level atom interacting with a massless scalar field, where they evaluated the contributions of vacuum fluctuations and radiation reaction to both the spontaneous excitation and radiative energy shifts of the atom. This framework was further generalized to atoms on arbitrary stationary trajectories \cite{Audretsch1995CQG}, providing a unified treatment of the generalized Unruh effect and the Lamb shift for noninertial atoms. 
Years later, Passante \cite{Passante1998PRA} investigated the radiative level shifts of a more realistic system, a uniformly accelerated hydrogen atom, interacting with a quantized electromagnetic field using the DDC formalism.
Since then, the DDC formalism has been widely applied to investigate the Lamb shift of atoms in various situations, including atoms moving along different trajectories and coupled to different quantum fields \cite{Cai2022EPJC}, the consideration of boundary conditions on quantum fields \cite{Rizzuto2007PRA,Rizzuto2009PRA,Zhu2009PRA,Zhu2010PRA}, and extensions to curved spacetime \cite{Zhou2010PRD1,Zhou2010PRD2,Zhou2012JHEP,Bohra2021EPJC,Yu2023JHEP}. Collectively, these studies have established that the vacuum fluctuation and radiation reaction contributions to the relative radiative energy shift are sensitive probes of the global structure of spacetime and the state of the quantum field.

On a separate front, the existence of a minimum measurable length at the Planck scale is a generic prediction of various approaches to quantum gravity, including string theory \cite{Veneziano1986EPL,Amati1989PLB,Konishi1990PLB}, loop quantum gravity \cite{Rovelli1995NPB,Ashtekar2004CQG}, asymptotically safe gravity \cite{Reuter1998PRD,Niedermaier2006LRR}, and non-commutative geometry \cite{Connes1994Book,Douglas2001RMP}.
The minimal length leads to a generalization of the Heisenberg Uncertainty Principle (HUP) to the Generalized Uncertainty Principle (GUP), whose simplest form reads
\begin{align}
	\Delta X \Delta P \geq \frac{\hbar}{2} \left[ 1 + \beta (\Delta P)^2 \right],
\end{align}
where $\beta = \beta_0 \ell_{\mathrm{Pl}}^2 / \hbar^2$ is the GUP parameter, $\beta_0$ is a dimensionless constant of order unity, and $\ell_{\mathrm{Pl}} = \sqrt{G\hbar/c^3} \simeq 1.6 \times 10^{-35}$ m is the Planck length. The GUP was first suggested from string theory considerations by Amati, Ciafaloni, and Veneziano \cite{Amati1987PLB,Amati1989PLB}, building on the early insight that string theory requires only two fundamental constants \cite{Veneziano1986EPL}; it was independently derived from a gedanken experiment in black hole physics by Maggiore \cite{Maggiore1993PLB1}, and later generalized to deformed algebras \cite{Maggiore1993PLB2}. Scardigli \cite{Scardigli1999PLB} later provided an independent derivation based on micro-black hole gedanken experiments. Over the past three decades, a variety of GUP models have been proposed, differing in their algebraic structure and the form of the modified commutation relations. The most widely used model is that of Kempf, Mangano, and Mann (KMM) \cite{Kempf1995PRD}, which implements the minimal length through a deformation of the canonical commutation relations. The quadratic form (1) is the hallmark of the KMM model. Extensions of the KMM model include higher-order GUP formulations with both minimal length and maximal momentum \cite{Pedram2012PLB1,Pedram2012PLB2}, non-perturbative GUP constructions \cite{Jizba2010PRD}, and GUP models incorporating both linear and quadratic momentum corrections \cite{Das2010PLB,Ali2009PLB,Majumder2011PLB}.
For comprehensive reviews, we refer to Refs. \cite{Hossenfelder2013LRR,Tawfik2014IJMPD,Tawfik2015RPP,Bosso2023CQG}.

The GUP can modify the Hamiltonian of classical and quantum systems in a general manner \cite{Das2011PLB,Das2008PRL}, thereby giving rise to quantum gravitational corrections to a wide range of physical phenomena. A large body of recent research has been devoted to investigating the effects of the GUP on such classical and quantum systems \cite{Benczik2005PRA,Bang2006PRD,Nozari2010EPL,Bawaj2015NatCommun,Nicolini2011PLB,Husain2016PRL,Gim2018PLB,Scardigli2018EPJC,Marin2013NatPhys,Bosso2017PRA,Bosso2017PRD,Kumar2018PRA,Mania2011PLB,Berger2011PRD,Pedram2011JHEP,Haouat2014PLB,Bernardo2016AnnPhys,Villalpando2019PRD,Vakili2012JSM,AntonacciOakes2013EPJC,Guo2016JCAP,Samar2016JMP,Castro2017JPA,Prasetyo2022EPJC,Twagirayezu2020AnnPhys,Petruzziello2021NatCommun,Fadel2022PRD,Bosso2023CommPhys,Gomes2023CQG,Ali2022PLB,Casadio2023PLB,Artigas2024PRD,Ali2011PRD,Gao2016PRA,AmelinoCamelia2009PRL,Bushev2019PRD,Pikovski2012NatPhys,Kumar2020NatCommun,Sen2022CQG}. In quantum field theory, the GUP has been shown to modify the Casimir effect \cite{Harbach2006PLB,Frassino2012PRD} and the Unruh effect and the Unruh temperature \cite{Scardigli1995NCB,Majhi2013PLB}. In black hole physics, the GUP has been extensively applied to modify the Hawking temperature and the thermodynamics of black holes \cite{Adler2001GRG,Arzano2005JHEP}, to induce logarithmic corrections to the Bekenstein-Hawking entropy \cite{Medved2004PRD}, to alter the evaporation process and leave behind a Planck-scale remnant \cite{Ong2018JHEP}, and to shift the quasinormal mode spectrum of black holes \cite{Das2021PLB,Jusufi2022Universe}. In cosmology, GUP corrections have been applied to the Friedmann equations and the early Universe \cite{Salah2017JCAP,Awad2014JHEP}.

A particularly promising feature of GUP phenomenology is that, although the fundamental scale is the Planck length, the resulting corrections may be amplified by macroscopic parameters, notably by the proper acceleration of a noninertial detector. 
Wang \cite{Wang2025NPB} applied the DDC approach to study the GUP corrections to the spontaneous radiation properties of a two-level atom interacting with a GUP-modified massless scalar field. The key finding was that the GUP alters the correlation function of the scalar field and consequently modifies both the spontaneous emission rate of an inertial excited atom and the transition rates of an accelerated atom. Notably, the GUP-induced corrections were shown to contain terms proportional to $a^2$ or $a^3$, confirming that the proper acceleration can significantly amplify the quantum gravity signal. The GUP was also found not to alter the stability of the ground-state atom in vacuum. Wang and Long \cite{Wang2025PLB} subsequently extended this analysis to a multilevel atom coupled to GUP-modified electromagnetic vacuum fluctuations. For the electromagnetic case, the GUP corrections to transition rates were found to contain components proportional to $a^4$, and the sign of the GUP parameter was shown to determine whether the transition rates are enhanced or suppressed relative to the standard case.

Despite these important advances in understanding the GUP corrections to atomic transition rates, a notable gap remains. Within the DDC formalism, the vacuum fluctuation and radiation reaction contributions govern not only the rate of change of the atomic energy but also the shift of the atomic energy levels themselves, i.e., the Lamb shift. For an accelerated atom, the Lamb shift receives distinct contributions from vacuum fluctuations and radiation reaction, and its dependence on the atomic acceleration, the spacetime curvature, and the presence of boundaries has been systematically studied in a variety of contexts. The Lamb shift is fundamentally different from transition rates, since the vacuum fluctuation and radiation reaction contributions to the Lamb shift and to the transition rates are governed by different combinations of the field correlation functions, and can therefore exhibit qualitatively distinct behaviors. To our knowledge, the GUP corrections to the Lamb shift of an accelerated atom have not been investigated within the DDC formalism.

In this paper, we present the first systematic study of the GUP corrections to the Lamb shift of a two-level atom undergoing inertial motion, uniform acceleration, and uniform circular motion, interacting with a massless quantum scalar field, within the DDC formalism. We adopt the KMM GUP model \cite{Kempf1995PRD} to introduce the GUP-modified positive frequency Wightman function of the scalar field. Within the DDC framework, we separate the contributions of vacuum fluctuations and radiation reaction to the Lamb shift, evaluating each contribution for both inertial and noninertial atoms. We find that the GUP yields nontrivial corrections to the Lamb shift that depend on both the GUP parameter $\beta$ and the proper acceleration $a$. The acceleration-dependent terms exhibit a form distinct from those found in the transition rate corrections of Refs. \cite{Wang2025NPB,Wang2025PLB}, reflecting the qualitatively different nature of the Lamb shift as an observable. This distinction may provide an independent channel for probing quantum gravity effects at low energies.

The paper is organized as follows. In Sec. II, we brief review 
the GUP-modified scalar field Wightman function from the KMM model and
the essential elements of the DDC formalism, establishing the basic formulas for the vacuum fluctuation and radiation reaction contributions to the Lamb shift. In Sec. III, we calculate the GUP corrections to the Lamb shift for an inertial two-level atom and verify that the inertial Lamb shift reduces to the known result in the limit $\beta \to 0$. 
In Secs. IV and V, we extend this discussion to the Lamb shift of atoms in noninertial motion. We provide a detailed analysis of the acceleration-dependent correction induced by the GUP as a function of the proper acceleration of the atom, for uniformly accelerated motion and uniform circular motion, respectively.
We conclude with a summary in last Section. Starting from the next section, we work in natural units $\hbar = c = k_B = 1$.

\section{The general formalism}

\subsection{The GUP-modified positive frequency Wightman function}

We start from the KMM GUP model \cite{Kempf1995PRD}, which is given by Eq. (1). At energies much lower than the Planck energy, the correction caused by the GUP becomes negligible and the HUP will be recovered.
It is easy to see that the uncertainty relation (1) corresponds to a minimum position uncertainty. For mirror-symmetric states, the uncertainty relation (1) can be derived straightforwardly by utilizing the commutator $[X, P] = i (1 + \beta P^2)$. In the three-dimensional scenario, the most general rotationally invariant expression is $[X_i, P_j] = i\left(\delta_{ij} + \beta P^2 \delta_{ij} + \beta' P_i P_j\right)$, where, as frequently done in the literature [41,42], we take $\beta' = 2\beta$. At linear order in $\beta$, this algebra is realized by the conventional position and momentum operators,
$X_i = x_i, P_i = p_i\left(1 + \beta \mathbf{p}^2\right),$
with $[x_i, p_j] = i\delta_{ij}$. Due to quantum gravitational fluctuations, the momentum $\mathbf{p}$ gains an increment of $\beta \mathbf{p}^2 \mathbf{p}$, resulting in a modified dispersion relation
\cite{Mania2011PLB,Berger2011PRD}
\begin{align}
E^{2}=\mathbf{p}^{2}+m^{2}+2 \beta \mathbf{p}^{4} .
\end{align}

Here a few comments are in order.
First, a number of studies in recent years have proposed that the parameter associated with the GUP may actually be negative rather than positive as is commonly assumed in certain scenarios. 
For example, Ong shows in the analysis of the Chandrasekhar limit that $\beta$ should be negative in order to have a match between the white dwarfs masses predicted by GUP and the astrophysical measures \cite{Ong}. Moreover, the research referred to the Magueijo and Smolin formulation of doubly special relativity \cite{Magueijo2002PRL,Magueijo2005PRD}, the uncertainty relation from a crystal-like universe \cite{Jizba2010PRD}, the sparsity of Hawking radiation affected by rainbow gravity \cite{Feng}, also predict negative parameter.
As a result, various quantum gravitational models may yield differing signs for this parameter (whether positive or negative), suggesting that the sign of the GUP parameter should be regarded as a variable object \cite{Petruzziello,Du}.
Second, it is seen that the above dispersion relation explicitly violate Lorentz invariance.
The velocities of photons are energy-dependent due to the GUP, allowing for both subluminal and superluminal photon propagation. While superluminality may seem unphysical, as photons are expected to travel at the speed of light ($c$) in a vacuum according to Special Relativity, one can reasonably argue that the principle of relativity may not apply near or above the Planck scale $E_\mathrm{Pl}$ \cite{Hossenfelder2013LRR}. Furthermore, it has been demonstrated that superluminality can be interpreted as an apparent phenomenon within the framework of General Relativity. Specifically, photons may exhibit subluminal or superluminal behavior depending on the paths they take through a gravitational field and the observer's position \cite{Lust}.
Third, the form of Eq. (2) is also consistent with the effective field theory (EFT) approach to Lorentz violation, where the quadratic $p^4$ correction emerges as a theoretically well-motivated and self-consistent leading-order modification. Within the EFT framework, specifically the Standard-Model Extension (SME), the quadratic correction in Eq. (2) arises naturally from dimension-6 CPT-even operators. The general dispersion relation for photons in this framework takes the form $\omega^2 = k^2 + \xi_{\pm}^{(6)} k^4/E_{\text{Pl}}^2$ \cite{Shao2010MPLA}, which directly maps to Eq. (2) via the identification $2\beta = \xi^{(6)}/E_{\text{Pl}}^2$. The dimension-6 operators are the lowest-dimensional CPT-even Lorentz-violating operators in the EFT expansion, rendering the quadratic $p^4$ correction the most natural leading-order LV term that simultaneously preserves both CPT symmetry and helicity invariance, thereby avoiding the stringent birefringence constraints that would otherwise apply to lower-dimensional operators. The self-consistency of this quadratic form is further reinforced by its stable renormalization structure within the dimension-6 EFT, where the operators contributing to the $p^4$ term are protected by the residual symmetries of the SME framework. This is also reflected in the general parametrization $E^2 - p^2 = m^2 + \eta_{n} p^{n+2}/E_{\text{Pl}}^n$ \cite{Shao2010MPLA,Li2021PRD,Li2021PLB,Li2023JCAP}, which, for the $n=2$ case, precisely reproduces Eq. (2) and identifies the $p^4$ correction as a specific instance of the low-energy expansion of quantum spacetime. The overall theoretical consistency of the quadratic correction is evidenced by its emergence across multiple independent quantum gravity-motivated frameworks, this convergence suggests that the quadratic modification is probably a robust feature of the low-energy effective description of quantum spacetime.
In the limit $\beta \to 0$, the standard dispersion relation with no GUP correction is recovered.

The dispersion relation (2) leads to the GUP-modified propagator of
the scalar quantum field in position space as
\begin{align}
G\left(x, x^{\prime}\right)=\int_{}^{} \frac{\mathrm{d}^{4} p}{(2 \pi)^{4}} \frac{e^{-i\left[p_{0}\left(t-t^{\prime}\right)-\mathbf{p} \cdot\left(\mathbf{x}-\mathbf{x}^{\prime}\right)\right]}}{p_{0}^{2}-\mathbf{p}^{2}\left(1+2\beta \mathbf{p}^{2}\right)-m^{2}},
\end{align}
the $ p_{0} $ integral is performed by use of a contour integral, and the poles occur at $ p_{0}^{2}=\mathbf{p}^{2}\left(1+2\beta \mathbf{p}^{2}\right)+m^{2}$, with the contour selection relevant to the two-point functions in the standard approach \cite{Birrell}. 
Then the positive frequency Wightman function modified by the GUP in the massless limit can be obtained as \cite{Davies}
\begin{align}
	D^{+}\left(x, x^{\prime}\right) =-\frac{1}{4 \pi ^{2}} \frac{1}{(\Delta t-i \varepsilon)^{2}-|\Delta \mathbf{x}|^{2}}\left(1-\frac{2 \beta}{(\Delta t-i \varepsilon)^{2}-|\Delta \mathbf{x}|^{2}}\right),
\end{align}
where the spacetime points are $x = (t, \mathbf{x})$ and an infinitesimally small positive
parameter $\varepsilon$ is introduced to characterize the singularities of the function.

\subsection{Vacuum fluctuations and radiation reaction contributions to the radiative energy shifts}
Consider a point-like two-level atom weakly coupled to a quantized massless scalar field. The atom has two stationary eigenstates, $|+\rangle$ and $|-\rangle$, with energies $\frac{1}{2}\omega_0$ and $-\frac{1}{2}\omega_0$, respectively. The evolution of the atom-field system with respect to the atom’s proper time $\tau$ is governed by the Hamiltonian,
\begin{align}
	H (\tau ) = {H_A}(\tau ) + {H_F}(\tau ) + {H_I}(\tau ),
\end{align}
where $H_A(\tau)$ is the Hamiltonian of the atom, it is given by
${H_A}(\tau )=  {\omega _0}{R _3}(\tau )$
with
$R_3(0)=\frac{1}{2}|+\rangle\langle+|-\frac{1}{2}|-\rangle\langle-|$. $H_F(\tau)$ is the free Hamiltonian of the quantum scalar field
\begin{align}
	H_{F}(\tau)=\int d^{3} k \omega_{\vec{k}} a_{\vec{k}}^{\dagger} a_{\vec{k}} \frac{d t}{d \tau},
\end{align}
with \(a_{\vec{k}}^\dagger\) and \(a_{\vec{k}}\) being the creation and annihilation operators for momentum \(\vec{k}\). The interaction between the atom and the field is governed by  
$H_I(\tau) = \mu\, R_2(\tau)\,\phi(x(\tau))$, 
where \(\mu\) is a small coupling constant. The operator \(R_2(0)\) is defined as \(R_2(0) = \frac{1}{2}i[R_-(0) - R_+(0)]\), with the atomic raising and lowering operators given by \(R_+(0) = |+\rangle\langle-|\) and \(R_-(0) = |-\rangle\langle+|\). Together with \(R_3\), these operators satisfy the angular momentum algebra,  
$[R_3, R_\pm] = \pm R_\pm$, $ [R_+, R_-] = 2R_3 $.  
\(\phi(x)\) is the scalar field operator satisfying the Klein–Gordon equation. The coupling is effective only on the atomic trajectory \(x(\tau)\).

From the above Hamiltonians, one can derive the Heisenberg equations of motion for the atomic and field variables. Their solutions naturally separate into two parts: a free part that survives even in the absence of coupling, and a source part induced by the interaction. We assume that the atom is initially prepared in the state \(|b\rangle\) and the field is in the vacuum state \(|0\rangle\). To distinguish the contributions of vacuum fluctuations and radiation reaction to the energy-level shift of a two-level atom, we employ the DDC formalism, which prescribes a symmetric ordering between the atom and field variables. Proceeding in a manner similar to Refs. \cite{Audretsch1995PRA,Audretsch1995CQG}, we obtain the following effective Hamiltonians for the two contributions, valid up to order \(\mu^2\), as
\begin{align}
	H_{v f}^{\mathrm{eff}}(\tau) & = \frac{1}{2} i \mu^{2} \int_{\tau_{0}}^{\tau} d \tau^{\prime} C^{F}\left(x(\tau), x\left(\tau^{\prime}\right)\right)\left[R_{2}^{f}\left(\tau^{\prime}\right), R_{2}^{f}(\tau)\right]\\
	H_{r r}^{\mathrm{eff}}(\tau) & = -\frac{1}{2} i \mu^{2} \int_{\tau_{0}}^{\tau} d \tau^{\prime} \chi^{F}\left(x(\tau), x\left(\tau^{\prime}\right)\right)\left\{R_{2}^{f}\left(\tau^{\prime}\right), R_{2}^{f}(\tau)\right\},
\end{align}
Here $\{\;,\;\}$ and $[\;,\;]$ denote the commutator
and anticommutator separately. $C^{F}\left(x(\tau), x\left(\tau^{\prime}\right)\right)$ and $\chi^{F}\left(x(\tau), x\left(\tau^{\prime}\right)\right)$ represent the symmetric correlation function and linear susceptibility of the scalar field in the vacuum state $\left|0\right\rangle$, respectively,
defined as
\begin{align}
	C^{F}\left(x(\tau), x\left(\tau^{\prime}\right)\right) & =\frac{1}{2}\left\langle 0\left|\left\{\phi^{f}(x(\tau)), \phi^{f}\left(x\left(\tau^{\prime}\right)\right)\right\}\right| 0\right\rangle, \\
	\chi^{F}\left(x(\tau), x\left(\tau^{\prime}\right)\right) & =\frac{1}{2}\left\langle 0\left|\left[\phi^{f}(x(\tau)), \phi^{f}\left(x\left(\tau^{\prime}\right)\right)\right]\right| 0\right\rangle .
\end{align}
We thus obtain the contributions from vacuum field fluctuations and radiation reaction to the energy shift of level \(b\) by taking the expectation values of Eqs. (7) and (8) in a generic atomic state \(|b\rangle\), 
\begin{align}
	\left(\delta E_{b}\right)_{v f} & = -i \mu^{2} \int_{\tau_{0}}^{\tau} d \tau^{\prime} C^{F}\left(x(\tau), x\left(\tau^{\prime}\right)\right) \chi_{b}^{A}\left(\tau, \tau^{\prime}\right) \\
	\left(\delta E_{b}\right)_{r r} & = -i \mu^{2} \int_{\tau_{0}}^{\tau} d \tau^{\prime} \chi^{F}\left(x(\tau), x\left(\tau^{\prime}\right)\right) C_{b}^{A}\left(\tau, \tau^{\prime}\right)
\end{align}
where \(C^A_b(\tau,\tau')\) and \(\chi^A_b(\tau,\tau')\) denote the symmetric correlation function and the linear susceptibility of the atom, respectively, and are given by
\begin{align}
	C^{A}\left(\tau, \tau^{\prime}\right) & =\frac{1}{2}\left\langle b\left|\left\{R_{2}^{f}(\tau), R_{2}^{f}\left(\tau^{\prime}\right)\right\}\right| b\right\rangle, \\
	\chi^{A}\left(\tau, \tau^{\prime}\right) & =\frac{1}{2}\left\langle b\left|\left[R_{2}^{f}(\tau), R_{2}^{f}\left(\tau^{\prime}\right)\right]\right| b\right\rangle,
\end{align}
Both are independent of the atomic trajectory and are determined entirely by the intrinsic properties of the atom. Explicitly, they read
\begin{align}
	C_{b}^{A}\left(\tau, \tau^{\prime}\right) & =\frac{1}{2} \sum_{d}\left|\left\langle b\left|R_{2}^{f}(0)\right| d\right\rangle\right|^{2}\left(e^{i \omega_{b d}\Delta \tau}+e^{-i \omega_{b d}\Delta \tau}\right), \\
	\chi_{b}^{A}\left(\tau, \tau^{\prime}\right) & =\frac{1}{2} \sum_{d}\left|\left\langle b\left|R_{2}^{f}(0)\right| d\right\rangle\right|^{2}\left(e^{i \omega_{b d}\Delta \tau}-e^{-i \omega_{b d}\Delta \tau}\right),
\end{align}
where $\omega_{b d}=\omega_b-\omega_d$ and $\Delta \tau= \tau- \tau'$, the sum spreads over a complete set of atomic stationary states.

Plugging the positive frequency Wightman function (4), the statistical functions modified by the GUP of the field can be written as
\begin{equation}\begin{aligned}
		C^{F}\left(x, x^{\prime}\right) 
		= & -\frac{1}{8 \pi^{2}}\left(\frac{1}{(\Delta t-i \varepsilon)^{2}-|\Delta \mathbf{x}|^{2}}+\frac{1}{(\Delta t+i \varepsilon)^{2}-|\Delta \mathbf{x}|^{2}}\right)+ \\
		& \frac{\beta}{4 \pi^{2}}\left(\frac{1}{\left((\Delta t-i \varepsilon)^{2}-|\Delta \mathbf{x}|^{2}\right)^{2}}+\frac{1}{\left((\Delta t+i \varepsilon)^{2}-|\Delta \mathbf{x}|^{2}\right)^{2}}\right), 
\end{aligned}\end{equation}	
\begin{equation}\begin{aligned}
		\chi^{F}\left(x, x^{\prime}\right)= & -\frac{1}{8 \pi^{2}}\left(\frac{1}{(\Delta t-i \varepsilon)^{2}-|\Delta \mathbf{x}|^{2}}-\frac{1}{(\Delta t+i \varepsilon)^{2}-|\Delta \mathbf{x}|^{2}}\right)+ \\
		& \frac{\beta}{4 \pi^{2}}\left(\frac{1}{\left((\Delta t-i \varepsilon)^{2}-|\Delta \mathbf{x}|^{2}\right)^{2}}-\frac{1}{\left((\Delta t+i \varepsilon)^{2}-|\Delta \mathbf{x}|^{2}\right)^{2}}\right),
\end{aligned}\end{equation}
where $\Delta t = t(\tau ) - t(\tau '),\Delta \mathbf{x} = \mathbf{x}(\tau ) - \mathbf{x}(\tau ')$.

\section{The uniformly moving atom}
Using the DDC formalism introduced above, we investigate in this section the GUP-induced corrections to the Lamb shift of an inertial atom. The results will serve as a basis for the analysis of vacuum fluctuations and radiation reaction contributions modified by the GUP in the more general case of an accelerating atom, treated in the subsequent sections. Specifically, we take an inertial atom traveling along the \(x\)-direction at a constant velocity \(v\), then the trajectory of the atom reads
\begin{align}
t(\tau)=\gamma \tau, \quad \mathbf{x}(\tau)=\mathbf{x}_{0}+\mathbf{v} \gamma \tau,
\end{align}
where the Lorentz
factor $\gamma=(1-\mathbf{v}^2)^{-1/2}$. 
The statistical functions of the field, which follow straightforwardly from Eqs. (17) and (18), are best expressed for our purposes as frequency integrals
\begin{equation}\begin{aligned}
		C^{F}\left(x, x^{\prime}\right) = \frac{1}{8\pi^{2}}\int_{0}^{\infty} d\omega\omega(e^{-i\omega\Delta\tau}+e^{i\omega\Delta\tau}) 
		+ \frac{\beta}{24 \pi^{2}} \int_{0}^{\infty}d\omega\omega^{3}(e^{-i\omega\Delta\tau}+e^{i\omega\Delta\tau}),
\end{aligned}\end{equation}	
\begin{equation}\begin{aligned}
		\chi^{F}\left(x, x^{\prime}\right) = \frac{1}{8\pi^{2}}\int_{0}^{\infty} d\omega\omega(e^{-i\omega\Delta\tau}-e^{i\omega\Delta\tau})
		+ \frac{\beta}{24 \pi^{2}} \int_{0}^{\infty}d\omega\omega^{3}(e^{-i\omega\Delta\tau}-e^{i\omega\Delta\tau}).
\end{aligned}\end{equation}
With Eqs. (20) and (21), we can directly derive the vacuum field fluctuation and radiation reaction contributions to the average rate of change of atomic excitation energy, which were previously discussed in Ref. \cite{Wang2025NPB}.
Inserting the Eqs. (15), (16), (20) and (21) into Eqs. (11) and (12), and extending the range of integration to infinity for sufficiently long time interval, we obtain the contribution of vacuum field fluctuations and radiation reaction to the energy shift of the state $|b\rangle$ as
\begin{align}
	{\left\langle {\delta E_b} \right\rangle _{vf}} = \frac{{{\mu ^2}}}{{8{\pi ^2}}}\sum\limits_d {{{\left| {\langle b|R_2^f(0)|d\rangle } \right|}^2}\int d \omega \omega } \left( {1 + \frac{{\beta {\omega ^2}}}{3}} \right){\cal P}\left( {\frac{1}{{\omega  + {\omega _{bd}}}} - \frac{1}{{\omega  - {\omega _{bd}}}}} \right),
\end{align}
\begin{align}
	{\left\langle {\delta E_b} \right\rangle _{rr}} = - \frac{{{\mu ^2}}}{{8{\pi ^2}}}\sum\limits_d {{{\left| {\langle b|R_2^f(0)|d\rangle } \right|}^2}\int d \omega \omega } \left( {1 + \frac{{\beta {\omega ^2}}}{3}} \right){\cal P}\left( {\frac{1}{{\omega  + {\omega _{bd}}}} + \frac{1}{{\omega  - {\omega _{bd}}}}} \right).
\end{align}

As shown, the two contributions both exhibit a dependence on the GUP parameter \(\beta\).
For a two-level atom with an energy gap \(\omega_0\) between the ground state \(|-\rangle\) and the excited state \(|+\rangle\), the vacuum fluctuations contribution to the energy shifts of both levels can be readily obtained
\begin{equation}\begin{aligned}
		\left(\delta E_{+}\right)_{v f} =-\left(\delta E_{-}\right)_{v f} = \frac{\mu^{2}}{32 \pi^{2}} \int_{0}^{\infty} d \omega \omega \left( {1 + \frac{{\beta {\omega ^2}}}{3}} \right) \mathcal{P} \left(\frac{1}{\omega+\omega_{0}} - \frac{1}{\omega -\omega_{0}}\right),
\end{aligned}\end{equation}
where the relation $\sum_{d}\left|\left\langle b\left|R_{2}^{f}(0)\right| d\right\rangle\right|^{2}= 1/4$ has been employed. We find that the vacuum fluctuation contributions to the energy shifts of the two different levels differ only in sign, a feature that is the same as in the absence of GUP effect. For the contribution of the radiation reaction, we have
\begin{equation}\begin{aligned}
		\left(\delta E_{+}\right)_{rr} = \left(\delta E_{-}\right)_{rr} = - \frac{\mu^{2}}{32 \pi^{2}} \int_{0}^{\infty} d \omega \omega \left( {1 + \frac{{\beta {\omega ^2}}}{3}} \right) \mathcal{P} \left(\frac{1}{\omega+\omega_{0}} + \frac{1}{\omega -\omega_{0}}\right),
\end{aligned}\end{equation}
obviously, the radiation reaction contributes the same to each energy level shift, which is similar with the case without the GUP. Adding up the contributions of vacuum fluctuations and radiation reaction, we obtain the total energy shifts for each level
\begin{equation}\begin{aligned}
		\delta E_{+} &= \left(\delta E_{+}\right)_{vf} + \left(\delta E_{+}\right)_{rr}= - \frac{\mu^{2}}{16 \pi^{2}} \int_{0}^{\infty} d \omega \omega \left( {1 + \frac{{\beta {\omega ^2}}}{3}} \right) \mathcal{P} \frac{1}{\omega -\omega_{0}},\\
		\delta E_{-} &= \left(\delta E_{-}\right)_{vf} + \left(\delta E_{-}\right)_{rr}= - \frac{\mu^{2}}{16 \pi^{2}} \int_{0}^{\infty} d \omega \omega \left( {1 + \frac{{\beta {\omega ^2}}}{3}} \right) \mathcal{P} \frac{1}{\omega +\omega_{0}}.
\end{aligned}\end{equation}

Thus the GUP-modified Lamb shift as the relative radiative energy shift of the two-level atom is given by
\begin{equation}\begin{aligned}
		\Delta =\delta E_{+}-\delta E_{-}= \frac{\mu^{2}}{16 \pi^{2}} \int_{0}^{\infty} d \omega \omega \left( {1 + \frac{{\beta {\omega ^2}}}{3}} \right) \mathcal{P} \left(\frac{1}{\omega+\omega_{0}} - \frac{1}{\omega -\omega_{0}}\right).
\end{aligned}\end{equation}

We observe that, regardless of whether the GUP effect is considered, the radiation reaction makes no contribution to the relative energy shift of the two-level atom, because it contributes equally to both levels.
So the relative shift of the atomic energy is entirely caused by vacuum fluctuations and can be actually calculated directly by $\Delta=(\delta E_{+})_{vf}-(\delta E_{-})_{vf}$.
The $\beta$-dependent terms in Eq. (27) give the corrections due to the GUP. As a result, the relative radiative energy shift may be enhanced or weakened, depending on whether the GUP parameter $\beta$ is positive or negative.
In the limit $\beta\rightarrow 0$, which describes the case without the GUP corrections, then the relative energy shift of the two-level atom reduces to
\begin{equation}\begin{aligned}
		\Delta_0 = \frac{\mu^{2}}{16 \pi^{2}} \int_{0}^{\infty} d \omega \omega \mathcal{P} \left(\frac{1}{\omega+\omega_{0}} - \frac{1}{\omega -\omega_{0}}\right),
\end{aligned}\end{equation}
this is precisely the result in free Minkowski spacetime \cite{Audretsch1995PRA}.

\section{The uniformly accelerated atom}

We now generalize the analysis of the preceding section to a uniformly accelerated atom, and investigate the GUP-induced modifications to the Lamb shift for an atom coupled to a massless scalar field. Taking the atom to accelerate uniformly along the \(z\)-direction with proper acceleration \(a\), we can describe its trajectory as
\begin{align}
t(\tau)=\frac{1}{a} \sinh a \tau, \quad z(\tau)=\frac{1}{a} \cosh a \tau, \quad x(\tau)=y(\tau)=0.
\end{align}

Inserting (29) into the general expressions of the GUP-modified symmetric correlation function (17) and linear
susceptibility function (18) of the scalar field, and after some calculations, we obtain
\begin{equation}\begin{aligned}
	C^{F}\left(x(\tau), x\left(\tau^{\prime}\right)\right) &=-\frac{a^{2}}{32 \pi^{2}}\left[\frac{1}{\sinh ^{2}\left(\frac{a}{2}\left(\tau-\tau^{\prime}\right)-i a \epsilon\right)}+\frac{1}{\sinh ^{2}\left(\frac{a}{2}\left(\tau-\tau^{\prime}\right)+i a \epsilon\right)}\right] \\
	&+\frac{\beta a^{4}}{64 \pi^{2}} \left[\frac{1}{\sinh ^{4}\left(\frac{a}{2}\left(\tau-\tau^{\prime}\right)-i a \epsilon\right)}+\frac{1}{\sinh ^{4}\left(\frac{a}{2}\left(\tau-\tau^{\prime}\right)+i a \epsilon\right)}\right], \\
\end{aligned}\end{equation}	
\begin{equation}\begin{aligned}	
	\chi^{F}\left(x(\tau), x\left(\tau^{\prime}\right)\right) &=-\frac{a^{2}}{32 \pi^{2}}\left[\frac{1}{\sinh ^{2}\left(\frac{a}{2}\left(\tau-\tau^{\prime}\right)-i a \epsilon\right)}-\frac{1}{\sinh ^{2}\left(\frac{a}{2}\left(\tau-\tau^{\prime}\right)+i a \epsilon\right)}\right] \\
	&+\frac{\beta a^{4}}{64 \pi^{2}} \left[\frac{1}{\sinh ^{4}\left(\frac{a}{2}\left(\tau-\tau^{\prime}\right)-i a \epsilon\right)}-\frac{1}{\sinh ^{4}\left(\frac{a}{2}\left(\tau-\tau^{\prime}\right)+i a \epsilon\right)}\right].
\end{aligned}\end{equation}

It is convenient to cast the field statistical functions (30) and (31) in the form of frequency integrals
\begin{align}
		C^{F}\left(x, x^{\prime}\right) = &\frac1{8\pi^2}\int_0^\infty d\omega\omega\coth\left(\frac{\pi\omega}a\right)\left(e^{-i\omega\Delta\tau}+e^{i\omega\Delta\tau}\right) + \notag \\
		& \frac{\beta}{24 \pi^{2}} \int_0^{\infty}d\omega\omega^3\left(1+\frac{a^2}{\omega^2}\right)\coth\left(\frac{\pi\omega}a\right)\left(e^{-i\omega\Delta\tau}+e^{i\omega\Delta\tau}\right), \\
		\chi^{F}\left(x, x^{\prime}\right) = &\frac1{8\pi^2}\int_0^\infty d\omega\omega\left(e^{-i\omega\Delta\tau}-e^{i\omega\Delta\tau}\right)
		+ \notag \\
	    &\frac{\beta}{24 \pi^{2}} \int_0^{\infty}d\omega\omega^3\left(1+\frac{a^2}{\omega^2}\right)\left(e^{-i\omega\Delta\tau}-e^{i\omega\Delta\tau}\right).
\end{align}

We note that both statistical functions are modified by the GUP; in particular, the linear susceptibility (33) of the field becomes dependent on the proper acceleration of the atom due to the GUP. 
Incidentally, using the statistical functions (32) and (33), we can directly derive the contributions of vacuum field fluctuations and radiation reaction to the average rate of change of the atomic excitation energy, as well as the total average rate of change of the atomic energy modified by the GUP, presented in Ref. \cite{Wang2025NPB}.
Consequently, the contributions from vacuum fluctuations and radiation reaction to the energy shift of the state $|b\rangle$ read, respectively,
\begin{align}
	{\left\langle {\delta E_b} \right\rangle _{vf}} = &\frac{{{\mu ^2}}}{{8{\pi ^2}}}\sum\limits_d {{\left| {\langle b|R_2^f(0)|d\rangle } \right|}^2} \times \notag \\
		&\int_0^{\infty} d \omega \omega \left( {1 + \beta \frac{{{\omega ^2} + {a^2}}}{3}} \right)\coth \left( {\frac{{\pi \omega }}{a}} \right) {\cal P}\left( {\frac{1}{{\omega  + {\omega _{bd}}}} - \frac{1}{{\omega  - {\omega _{bd}}}}} \right), \\
	{\left\langle {\delta E_b} \right\rangle _{rr}} = &- \frac{{{\mu ^2}}}{{8{\pi ^2}}}\sum\limits_d {{\left| {\langle b|R_2^f(0)|d\rangle } \right|}^2} \times \notag \\
		&\int_0^{\infty} d \omega \omega \left( {1 + \beta \frac{{{\omega ^2} + {a^2}}}{3}} \right){\cal P}\left( {\frac{1}{{\omega  + {\omega _{bd}}}} + \frac{1}{{\omega  - {\omega _{bd}}}}} \right).
\end{align}

The contributions of vacuum fluctuations to the energy shift of each level are then found to be
\begin{equation}\begin{aligned}
		\left(\delta E_{+}\right)_{v f} &= -\left(\delta E_{-}\right)_{v f} \\
		&= \frac{\mu^{2}}{32 \pi^{2}} \int_{0}^{\infty} d \omega \omega \left( {1 + \beta \frac{{{\omega ^2} + {a^2}}}{3}} \right) \left(1+\frac{2}{\mathrm{e}^{2 \pi\omega / a}-1}\right)
		\mathcal{P} \left(\frac{1}{\omega+\omega_{0}}-\frac{1}{\omega -\omega_{0}}\right),
\end{aligned}\end{equation}
while those of radiation reaction take the form 
\begin{equation}\begin{aligned}
		\left(\delta E_{+}\right)_{rr} = \left(\delta E_{-}\right)_{rr} = - \frac{\mu^{2}}{32 \pi^{2}} \int_{0}^{\infty} d \omega \omega \left( {1 + \beta \frac{{{\omega ^2} + {a^2}}}{3}} \right) \mathcal{P} \left(\frac{1}{\omega+\omega_{0}}+\frac{1}{\omega -\omega_{0}}\right).
\end{aligned}\end{equation}
It is seen that, when the GUP effect is taken into account, both the contributions from vacuum fluctuations and radiation reaction introduce additional correction terms that involve the GUP parameter \(\beta\) and the acceleration \(a\). As in the case without GUP, the radiation reaction contributes equally to the shift of each level and therefore does not affect the relative shift of the two levels. 
However, one should notice that the exact cancellation of radiation reaction contributions to the relative energy shifts for the two-level atom might be coincidental. For a realistic multilevel atomic system, this
fortuitous cancellation may no longer hold \cite{Audretsch1995PRA,Passante1998PRA}.
Hence, the Lamb shift of the two-level atom, which is entirely due to vacuum fluctuations, reads
\begin{equation}\begin{aligned}
		\Delta &=(\delta E_{+})_{vf}-(\delta E_{-})_{vf} \\
		&= \frac{\mu^{2}}{16 \pi^{2}} \int_{0}^{\infty} d \omega \omega \left( {1 + \beta \frac{{{\omega ^2} + {a^2}}}{3}} \right) \left(1+\frac{2}{\mathrm{e}^{2 \pi\omega / a}-1}\right) \mathcal{P} \left(\frac{1}{\omega+\omega_{0}}-\frac{1}{\omega -\omega_{0}}\right).
\end{aligned}\end{equation}
We see that the \(\beta\)-dependent terms give the GUP-induced corrections, which 
is structurally similar with the Lamb shift of a uniformly accelerated two-level atom coupled to electromagnetic vacuum fluctuations,
containing both thermal and nonthermal components.
The thermal term is characterized by the Planckian factor $\frac{1}{{{e^{2\pi \omega/a}} - 1}}$ with the Unruh temperature $T=a/2\pi$.
The Lamb shift can be enhanced for a positive GUP parameter, but weakened for a negative one. 
Especially, the corrections related to \(\beta a^2\) suggest that the atom’s acceleration can effectively amplify the GUP effect on the Lamb shift.
As \(\beta \to 0\), we recover the relative energy shift in the absence of GUP in free Minkowski spacetime \cite{Audretsch1995PRA}. In the limit \(a \to 0\), the result reduces to the GUP-modified Lamb shift for an inertial two-level atom shown in Eq. (27).

By subtracting the the relative energy shift of the uniformly accelerated two-level atom that does not take into account the GUP effect from Eq. (38),
we obtain the correction purely induced by GUP as
\begin{align}
	\Delta^{\rm{GUP}} =\Delta_0^{\rm{GUP}} + \Delta_T^{\rm{GUP}} + \Delta_{a}^{\rm{GUP}}, 
\end{align}
with
\begin{align}
	\Delta_{0}^{\rm{GUP}} = \frac{\beta {\gamma _0}}{6\pi {\omega _0}} \int_{0}^{\infty} d \omega \omega^{3} \mathcal{P} \left(\frac{1}{\omega+\omega_{0}}-\frac{1}{\omega-\omega_{0}}\right),
\end{align}
\begin{align}
	\Delta_{T}^{\rm{GUP}} = \frac{\beta {\gamma _0}}{6\pi {\omega _0}} \int_{0}^{\infty} d \omega \omega^{3} \frac{2}{e^{2 \pi \omega / a}-1} \mathcal{P} \left(\frac{1}{\omega+\omega_{0}}-\frac{1}{\omega-\omega_{0}}\right),
\end{align}
\begin{align}
	\Delta_{a-NT}^{\rm{GUP}} = \frac{\beta {\gamma _0}}{6\pi {\omega _0}} \int_{0}^{\infty} d \omega \omega a^{2}
	\operatorname{coth}\left(\frac{\pi \omega}{a}\right) \mathcal{P} \left(\frac{1}{\omega+\omega_{0}}-\frac{1}{\omega-\omega_{0}}\right).
\end{align}
Here, for convenience of discussion, the GUP correction is divided into three components, the first of which, $\Delta_{0}^{\rm{GUP}}$, is similar to the case of an inertial two-level atom interacting with quantum electromagnetic field fluctuations. It exhibits ultraviolet divergence under nonrelativistic theoretical frameworks, which is a known limitation of nonrelativistic quantum electrodynamics treatments. To address this divergence, one can introduce a frequency cutoff factor to suppress high-energy contributions or employ fully relativistic quantum field theory where the divergence naturally cancels through renormalization techniques without requiring artificial cutoffs \cite{Audretsch1995PRA,Passante1998PRA,Zhou2010PRD2}.
$\Delta_{T}^{\rm{GUP}}$ represents the acceleration-induced thermal GUP correction, which is analogous to the thermal correction for an inertial atom immersed in a thermal bath at temperature $T= \frac{a}{2\pi}$. Meanwhile, $\Delta_{a-NT}^{\rm{GUP}}$ denotes the nonthermal GUP correction associated with the atom's acceleration, which also exhibits divergence. To regularize the divergence, we can implement a cutoff frequency on the upper integration limit.
Here we are mainly concerned with the acceleration-related thermal and nonthermal corrections caused by the GUP, the variation of acceleration-induced thermal correction $\Delta_{T}^{\rm{GUP}}$ and nonthermal correction $\Delta_{a-NT}^{\rm{GUP}}$ as a function of the proper acceleration of the atom are respectively shown in Fig. 1 and Fig. 2.
\begin{figure}[htb]
	\centering
	\includegraphics[width=1.0\linewidth,angle=0,clip=true]{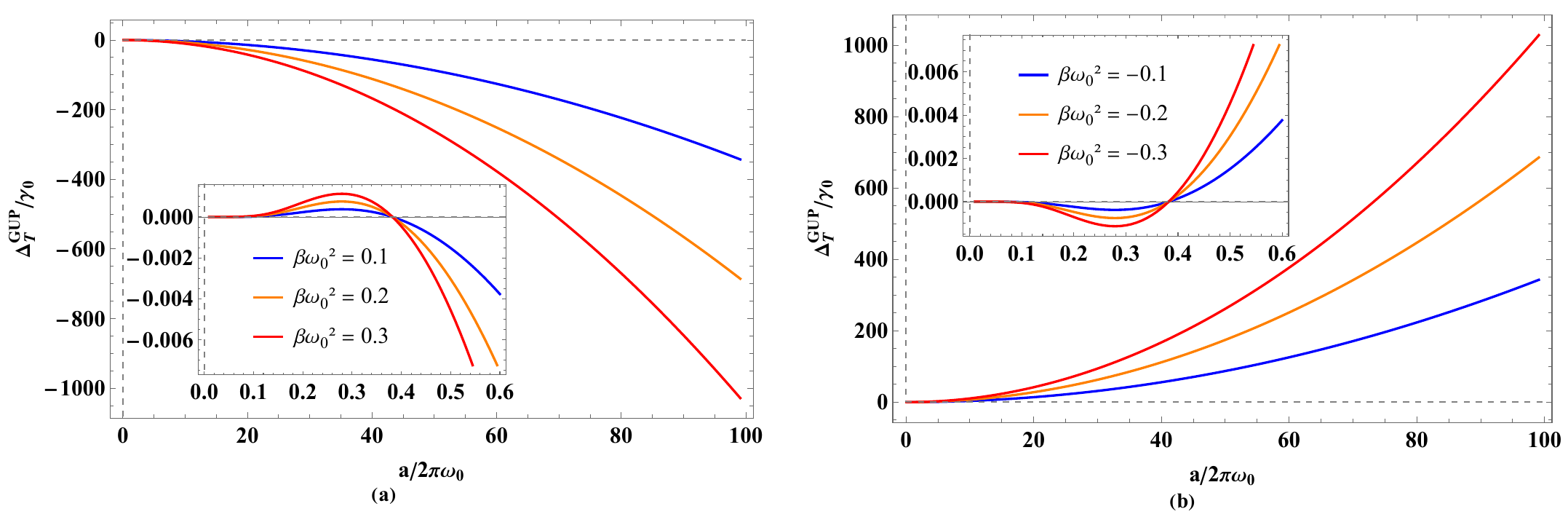}
	\caption{The acceleration-induced thermal GUP correction as a function of the atomic acceleration, with ${\gamma _0} = \frac{{{\mu ^2}{\omega _0}}}{{8\pi }}$ being the spontaneous emission rate of an inertial two-level atom interacting with a quantum scalar field in Minkowski vacuum. The left (a) and right (b) panels correspond to the cases of positive and negative GUP parameters, respectively.}
\end{figure}

We observed from Fig. 1 that, for the positive GUP parameter $\beta$, when the acceleration $a$ is relatively small, the acceleration induces a positive thermal correction. This correction initially increases with growing acceleration, gradually reaches a peak value, and then diminishes as a continues to rise. The thermal correction vanishes when the $a/2\pi \omega_{0}$ approaches approximately 0.383. As a further increases beyond this threshold, it begins to generate a negative thermal correction, whose magnitude grows progressively with increasing $a$. While for a negative parameter $\beta$, the trend of the curves are exactly the opposite.
In the limit $a/\omega_{0} \to 0$, the thermal correction can be approximated by
$\Delta_{T}^{\rm{GUP}}=\frac{\beta a^4}{360\pi \omega_{0}^2} \gamma_{0}$.
This indicates that when an atom undergoing uniformly accelerated motion has a very small acceleration, a thermal GUP correction proportional to $a^4$ can arise, which may offer a possible experimental scheme to probe the GUP by measuring the Unruh effect in low-acceleration scenarios.

\begin{figure}[htb]
	\centering
	\includegraphics[width=1.0\linewidth,angle=0,clip=true]{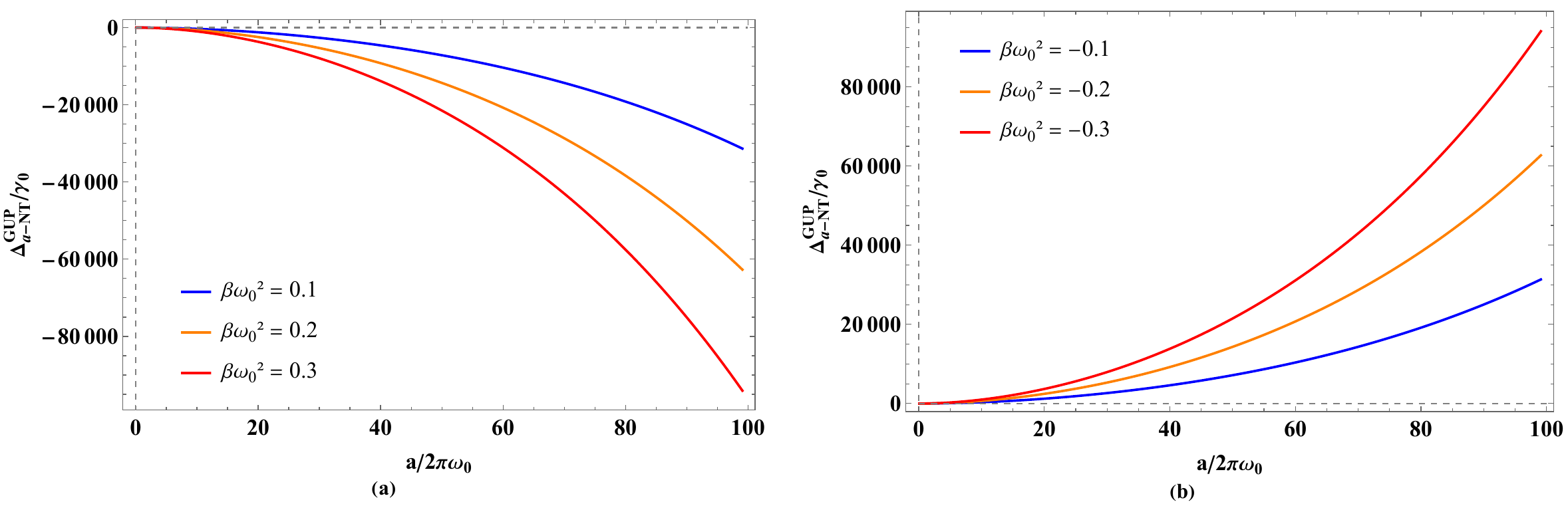}
	\caption{The acceleration-related nonthermal GUP correction as a function of the acceleration for a uniformly accelerated atom.}
\end{figure}
As shown in Fig. 2, the acceleration-dependent nonthermal corrections induced by the GUP exhibit a nonlinear monotonic behavior, increasing or decreasing sharply with acceleration depending on the sign of the parameter $\beta$. This implies that atomic acceleration acts as a powerful amplifier of quantum gravity effects, where microscopic Planck-scale corrections are increasingly magnified as acceleration grows, even for weak GUP couplings. For a fixed acceleration, the correction strength is directly proportional to the absolute magnitude of the GUP parameter, indicating that despite the smallness of $\beta$, detectable deviations can arise at sufficiently high accelerations. Therefore, for atoms undergoing large uniform acceleration, the nonthermal component of the Lamb shift can also serve as a sensitive probe of the GUP, providing additional experimental avenues for future tests of quantum gravity phenomenology.

\section{The atom in circular motion}
In this section, we turn to study the effect of GUP on the Lamb shift of a uniformly circulating atom of which the trajectory is described by
\begin{align}
	x(\tau ) = \left( {\gamma \tau ,R\cos (\gamma \Omega \tau ),R\sin (\gamma \Omega \tau ), 0} \right),
\end{align}
where $R $ is the radius of the orbit, $ \Omega $ is the angular velocity in the preferred Lorentz frame. 
and $\gamma=(1-v^2)^{-1/2}$ is the Lorentz factor with $v= R \Omega$.
The proper acceleration $ a = R \Omega^{2} \gamma^{2} $.

Inserting the (43) into Eqs. (17) and (18), in the ultrarelativistic limit $\gamma  \gg 1$, we obtain
the GUP-modified symmetric correlation function and the linear susceptibility of the field along the atomic 
trajectory, in terms of frequency integrations as
\begin{align}
	C^{F}\left(x, x^{\prime}\right) = &\frac{1}{8 \pi^{2}} \int_{0}^{\infty}d \omega \left(\omega+\frac{\sqrt{3} a}{6} e^{-\frac{2 \sqrt{3}}{a} \omega}\right)\left(e^{-i \omega \Delta\tau}+ e^{i \omega \Delta\tau}\right) +\notag \\ 
	& \frac{\beta}{24\pi^{2}}\int_{0}^{\infty}d\omega \left(\omega^{3}+a^{2}\omega+\frac{a^{2}}{4}\omega e^{-\frac{2\sqrt{3}}{a}\omega}+\frac{5 a^{3}}{8 \sqrt{3}} e^{-\frac{2 \sqrt{3}}{a}\omega}\right) 
	\left(e^{-i \omega \Delta\tau}+e^{i \omega \Delta\tau}\right),\\		
	\chi^{F}\left(x, x^{\prime}\right) = &\frac{1}{8 \pi^{2}} \int_{0}^{\infty}d \omega \omega \left(e^{-i \omega \Delta\tau}- e^{i \omega \Delta\tau}\right) +\notag \\
	& \frac{\beta}{24\pi^{2}}\int_{0}^{\infty}d\omega \left(\omega^{3}+a^{2}\omega\right)\left(e^{-i \omega \Delta\tau}-e^{i \omega \Delta\tau}\right).
\end{align}
We see both the symmetric correlation function (44) and the linear susceptibility (45) of the field are significantly modified by the GUP through the addition of several terms proportional to \(\beta\). In particular, we notice that the linear susceptibility of the field along the trajectory of an atom in circular motion is exactly the same as that for a uniformly accelerated atom, and thus also becomes dependent on the proper acceleration of the atom due to the GUP. 
Moreover, with the statistical functions (44) and (45) given above, we can directly derive the GUP-modified average rate of change of the atomic energy of a uniformly circulating atom, which matches the result reported in Ref. \cite{Wang2025NPB}.
The remaining calculation is straightforward. Inserting the statistical functions (15), (16), (44), and (45) into the general formulas (11) and (12) for the level shifts yields
\begin{equation}\begin{aligned}
	{\left\langle {\delta E_b} \right\rangle _{vf}} = &\frac{{{\mu ^2}}}{{8{\pi ^2}}}\sum\limits_d {{\left| {\langle b|R_2^f(0)|d\rangle } \right|}^2}\int d \omega \left[ {\omega  + \frac{{\sqrt 3 a}}{6}{e^{ - \frac{{2\sqrt 3 }}{a}\omega }} 
	 }\right.\\ & \left.{	
	+ \frac{\beta }{3}\left( {{\omega ^3} + {a^2}\omega  + \frac{{{a^2}}}{4}\omega {e^{ - \frac{{2\sqrt 3 }}{a}\omega }} + \frac{{5{a^3}}}{{8\sqrt 3 }}{e^{ - \frac{{2\sqrt 3 }}{a}\omega }}} \right)} \right] {\cal P}\left( {\frac{1}{{\omega  + {\omega _{bd}}}} - \frac{1}{{\omega  - {\omega _{bd}}}}} \right),
\end{aligned}\end{equation}
\begin{equation}\begin{aligned}
	{\left\langle {\delta E_b} \right\rangle _{rr}} = - \frac{{{\mu ^2}}}{{8{\pi ^2}}}\sum\limits_d {{{\left| {\langle b|R_2^f(0)|d\rangle } \right|}^2}\int d \omega \omega } \left( {1 + \beta \frac{{{\omega ^2} + {a^2}}}{3}} \right) {\cal P}\left( {\frac{1}{{\omega  + {\omega _{bd}}}} + \frac{1}{{\omega  - {\omega _{bd}}}}} \right).
\end{aligned}\end{equation}
Clearly, the contributions of vacuum field fluctuations and radiation reaction to each level are both dependent on the GUP parameter $\beta$ and the acceleration $a$ of the atom.
Once again, we can straightforwardly obtain contributions of vacuum field fluctuations to the energy shifts of the two levels
\begin{equation}\begin{aligned}
		\left(\delta E_{+}\right)_{v f} = &-\left(\delta E_{-}\right)_{v f} \\
		= &\frac{\mu^{2}}{32 \pi^{2}} \int_{0}^{\infty} d \omega \left[ {\omega  + \frac{{\sqrt 3 a}}{6}{e^{ - \frac{{2\sqrt 3 }}{a}\omega }}	
		+ \frac{\beta }{3}\left( {{\omega ^3} + {a^2}\omega  + \frac{{{a^2}}}{4}\omega {e^{ - \frac{{2\sqrt 3 }}{a}\omega }} + \frac{{5{a^3}}}{{8\sqrt 3 }}{e^{ - \frac{{2\sqrt 3 }}{a}\omega }}} \right)} \right]\\ 
	   & \times\mathcal{P} \left(\frac{1}{\omega+\omega_{0}} - \frac{1}{\omega -\omega_{0}}\right),
\end{aligned}\end{equation}
for the contribution of radiation reaction, we have
\begin{equation}\begin{aligned}
		\left(\delta E_{+}\right)_{rr} = \left(\delta E_{-}\right)_{rr} = - \frac{\mu^{2}}{32 \pi^{2}} \int_{0}^{\infty} d \omega \omega \left( {1 + \beta \frac{{{\omega ^2} + {a^2}}}{3}} \right) \mathcal{P} \left(\frac{1}{\omega+\omega_{0}} + \frac{1}{\omega -\omega_{0}}\right).
\end{aligned}\end{equation}
We observe that the radiation reaction contribution for an atom in uniform circular motion is identical to that for a uniformly accelerated atom. This is because the linear susceptibility of the field has the same form along the trajectory for both circular motion and uniform acceleration. Moreover, the radiation reaction shifts each energy level by exactly the same amount. Therefore, although this common shift contains GUP corrections related to the atom's acceleration, it does not contribute to the relative energy shift between the two levels, $\Delta_{r r} =\left(\delta E_{+}\right)_{r r}-\left(\delta E_{-}\right)_{r r}=0$.
Consequently, the GUP-modified Lamb shift of the two-level atom in circular motion is caused entirely by vacuum fluctuations, and reads
\begin{equation}\begin{aligned}
		\Delta \equiv & \Delta _{v f} + \Delta _{rr} =\left(\delta E_{+}\right)_{v f} - \left(\delta E_{-}\right)_{v f} \\
		= &\frac{\mu^{2}}{16 \pi^{2}} \int_{0}^{\infty} d \omega \left[ {\omega  + \frac{{\sqrt 3 a}}{6}{e^{ - \frac{{2\sqrt 3 }}{a}\omega }}	
			+ \frac{\beta }{3}\left( {{\omega ^3} + {a^2}\omega  + \frac{{{a^2}}}{4}\omega {e^{ - \frac{{2\sqrt 3 }}{a}\omega }} + \frac{{5{a^3}}}{{8\sqrt 3 }}{e^{ - \frac{{2\sqrt 3 }}{a}\omega }}} \right)} \right]\\ 
		& \times\mathcal{P} \left(\frac{1}{\omega+\omega_{0}} - \frac{1}{\omega -\omega_{0}}\right).
\end{aligned}\end{equation}

It is seen that the GUP-induced corrections consist of several \(\beta\)-dependent terms, which enhance or weaken the relative energy shifts of the uniformly circulating atom according to the sign of \(\beta\). In particular, the terms involving \(\beta a^2\) and \(\beta a^3\) indicate that the acceleration of the atom in uniform circular motion can also effectively amplify the GUP corrections to the Lamb shift. 
We notice that, in contrast to the linearly accelerated case, the Planckian terms appearing in the vacuum fluctuation contribution are replaced by terms with a non-Planckian exponential form in the circular motion case. This implies that the radiation experienced by an observer in uniform circular motion is intrinsically nonthermal. In the limit \(\beta \to 0\), the standard results in the absence of the GUP are recovered  \cite{Audretsch1995CQG}.
We further isolate the purely GUP-induced contributions to the Lamb shift as
\begin{align}
	\Delta^{\rm{GUP}} =\Delta_0^{\rm{GUP}} + \Delta_{a}^{\rm{GUP}}, 
\end{align}
where $\Delta_{0}^{\rm{GUP}}$ is the component that is structurally similar to the case of an inertial two-level atom interacting with fluctuations of the quantum electromagnetic field; its explicit expression is given in Eq. (40). $\Delta_{a}^{\rm{GUP}}$ is the acceleration-related GUP correction for a two-level atom in uniform circular motion, whose explicit expression reads
\begin{align}
	\Delta_{a}^{\rm{GUP}} = \frac{{\beta {\gamma _0}}}{{6\pi {\omega _0}}}\int_0^\infty  d \omega \left( {{a^2}\omega  + \frac{{{a^2}}}{4}\omega {e^{ - \frac{{2\sqrt 3 }}{a}\omega }} + \frac{{5{a^3}}}{{8\sqrt 3 }}{e^{ - \frac{{2\sqrt 3 }}{a}\omega }}} \right){\cal P}\left( {\frac{1}{{\omega  + {\omega _0}}} - \frac{1}{{\omega  - {\omega _0}}}} \right).
\end{align} 

In Fig. 3, we plot the acceleration-related correction induced by GUP as a function of atomic acceleration with different GUP parameter, for the case of a two-level atom in uniformly circular motion. The left (a) and right (b) panels correspond to the cases of positive and negative GUP parameters, respectively.
It is seen that, analogous to the case of uniform acceleration in linear motion, both display sharply nonlinear monotonic behaviors, undergoing a steep increase or decrease with increasing atomic acceleration, depending upon the positive or negative parameter $\beta$. For a fixed large acceleration, the corrections increase significantly with the absolute magnitude of the GUP parameter.
As the acceleration grows, the difference in the GUP-induced corrections for different GUP parameter values
becomes more pronounced. We also notice that the corrections for an atom in uniform circular motion are larger than those for a uniformly accelerated atom for the same $\beta$. This suggests that compared with uniformly accelerated linear motion, the acceleration for an atom in uniform circular motion more effectively amplifies the effect of GUP on the relative radiative energy shift. 
Thus, the Lamb shift of an atom undergoing centripetal acceleration may be more suitable as a sensitive probe of the GUP. Given that the extremely high acceleration required to observe GUP-induced corrections is challenging to achieve with linear motion, whereas large centripetal accelerations could be realized in certain settings, such as ultrarelativistic electrons in storage rings, this provides a potential avenue for experimentally testing GUP via precision measurements of the atomic Lamb shift.

\begin{figure}[htb]
	\centering
	\includegraphics[width=1.0\linewidth,angle=0,clip=true]{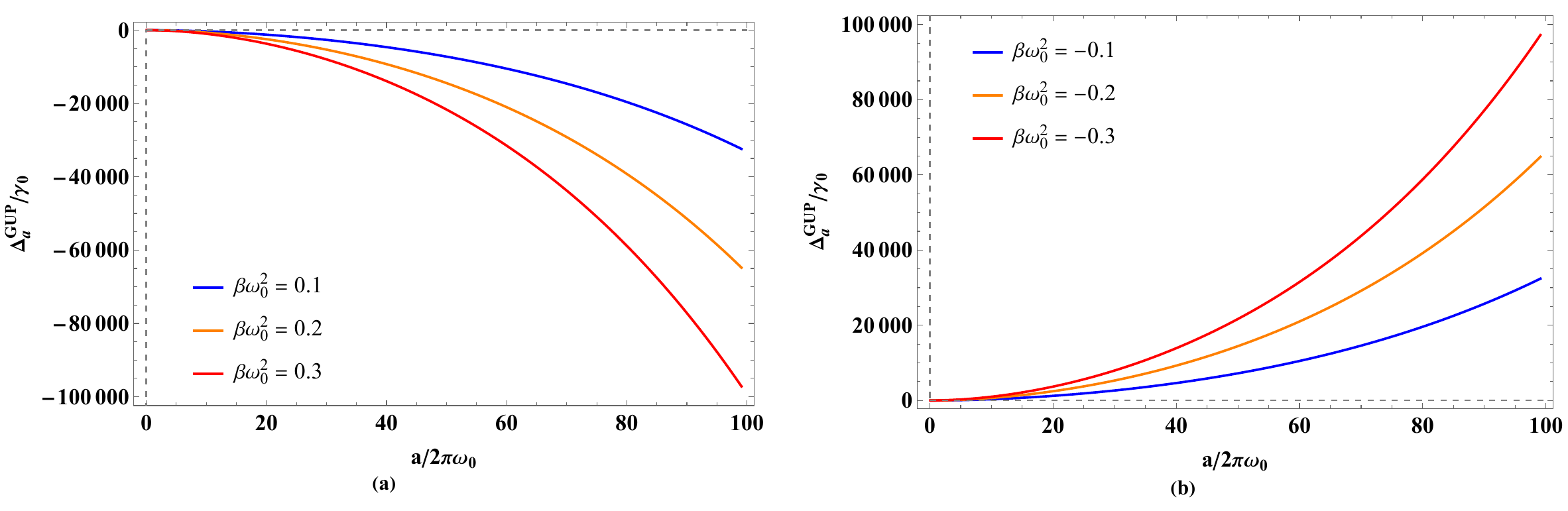}
	\caption{The acceleration-related GUP correction to the Lamb shift as a function of atomic acceleration, for the case of a two-level atom in uniformly circular motion.}
\end{figure}

\section{Summary}

The Generalized Uncertainty Principle (GUP) extends the Heisenberg Uncertainty Principle (HUP) by introducing a minimum observable length that encodes quantum gravitational effects. Although these effects are expected to become significant only near the Planck scale, they may leave observable imprints at much lower energies, offering a potential route to probing quantum gravity through low-energy physical processes.
In this paper, we investigate the effect of the GUP on the Lamb shift of a two-level atom interacting with a real massless scalar quantum field, and analyze the contributions of vacuum field fluctuations and radiation reaction for the atom in inertial motion, uniformly accelerated motion, and uniform circular motion, based on the DDC formalism.
We first calculate the statistical functions of the field along the trajectories for the three types of motion and express them as frequency integrals. It is found that both the symmetric correlation function and the linear susceptibility of the field acquire extra corrections proportional to $\beta$ due to the GUP. 
We then use these results to calculate the contributions of both vacuum field fluctuations and radiation reaction to the radiative level shift. 
It is also worth mentioning that, by expressing the statistical functions of the field as frequency integrals, we can directly reproduce the contributions of both vacuum field fluctuations and radiation reaction to the average rate of change of the atomic excitation energy reported in Ref. [13].

We find that for an inertial atom, both contributions to the energy shift of each level are modified by adding a term proportional to $\beta \omega^2$. For a uniformly accelerated atom, both contributions acquire a correction proportional to $\beta (\omega^2 + a^2)$. For a uniformly circulating atom, the radiation reaction contribution to each level is exactly the same as in the uniformly accelerated case, whereas the vacuum field fluctuations give rise to several additional terms involving both the GUP parameter $\beta$ and the acceleration $a$.
We observe that for atoms in all three types of motion, although the radiation reaction contributions are modified by the GUP, for the two-level atom considered here, the radiation reaction contributions to the relative energy shifts cancel out completely, because the radiation reaction contributes equally to the shift of each energy level. However, One should notice that the exact cancellation of radiation reaction contributions to the relative energy shifts for the two-level atom might be coincidental. For a realistic multilevel atomic system, this fortuitous cancellation may no longer hold [12,13].
Thus, the relative energy shift of the two-level atom arises entirely from vacuum fluctuations and can be calculated directly as  
$\Delta=(\delta E_{+})_{vf}-(\delta E_{-})_{vf}$. 
Consequently, the Lamb shift may be either enhanced or suppressed, depending on the sign of the GUP parameter \(\beta\). In the limit \(\beta\to 0\), the relative energy shift of the two-level atom for each type of motion reduces to the corresponding result in free Minkowski spacetime.

We focus on the analyze of the GUP correction to the Lamb shift of a two-level atom undergoing noninertial motion. 
It is seen for the uniformly accelerated two-level atom, the GUP-induced corrections, i.e., the purely $\beta$-dependent terms, which is structurally similar with the Lamb shift of a uniformly accelerated two-level atom coupled to electromagnetic vacuum fluctuations, containing both thermal with the usual Unruh temperature $T=a/2 \pi$ and nonthermal components. The corrections related to $\beta a^2$ suggest that the atom’s acceleration can effectively amplify the GUP effect on the Lamb shift. 
We also plot respectively the variation of acceleration-induced thermal correction $\Delta_{T}^{\rm{GUP}}$ and nonthermal correction $\Delta_{a-NT}^{\rm{GUP}}$ as a function of the proper acceleration of the atom with different GUP parameter in Fig. 1 and Fig. 2.
Overall, for a positive GUP parameter, the thermal correction exhibits nonmonotonic behavior at low accelerations, which first increases and then decreases, and eventually becomes monotonically decreasing as the acceleration increases further. For a fixed acceleration, increasing the GUP parameter enhances the magnitude of the correction. When the GUP parameter is negative, the trend of the curve is reversed. We also note that, at low accelerations, the thermal correction is approximately given by $\frac{\beta a^4}{360\pi \omega_{0}^2} \gamma_{0}$. This \(a^4\)-dependent thermal correction may provide a theoretical basis for probing the GUP through the Unruh effect at low accelerations.
By contrast, the nonthermal correction depends on acceleration in a purely nonlinear and monotonic manner. For a positive GUP parameter, it decreases monotonically with acceleration, while for a negative GUP parameter it increases monotonically. At large accelerations, increasing the GUP parameter significantly enhances the magnitude of the nonthermal correction. A comparison of Figs. 1 and 2 shows that the nonthermal correction is nearly two orders of magnitude larger than the thermal correction. This implies that, at large atomic accelerations, the nonthermal correction serves as a more sensitive probe of GUP effects than the thermal one.

For an atom in uniform circular motion, the GUP-induced corrections consist of several contributions. Unlike the uniformly accelerated case, there is no correction term involving the Planckian factor, which implies that the centripetal acceleration gives rise only to nonthermal GUP corrections. In particular, the terms proportional to \(\beta a^2\) and \(\beta a^3\) indicate that the acceleration of the atom in uniform circular motion can also effectively amplify the GUP corrections to the Lamb shift.
We further plot the GUP-induced corrections as functions of atomic acceleration for different values of \(\beta\). Similar to the nonthermal correction in the uniformly accelerated case, these corrections exhibit a strongly nonlinear and monotonic dependence on acceleration, increasing or decreasing steeply according to the sign of \(\beta\). At fixed acceleration, the correction magnitude increases significantly with \(|\beta|\). As the acceleration grows, the corrections corresponding to different values of the GUP parameter increasingly diverge. Comparing Figs. 2 and 3 reveals that, for the same \(\beta\), the corrections are larger for uniform circular motion than for uniformly accelerated motion, indicating that circular motion amplifies the GUP effect on the relative radiative energy shift more effectively than linear acceleration. The Lamb shift of an atom undergoing centripetal acceleration may therefore serve as a more sensitive probe of the GUP.

\section*{ACKNOWLEDGMENTS}
This work is supported by the Science and Technology Foundation of Guizhou Province (No. ZK[2022] General 029), the National Natural Science Foundation of China (No. 12505064), the Guizhou Provincial Basic Research Program (Natural Science) Youth Guidance Program (No. QN[2025]365), the Young Scientific and Technical Talents Development Project of the Education Department of Guizhou Province (No. [2024]79), and the Guizhou Provincial Basic Research Program General Project (No.~MS [2026] 068).

\end{document}